\documentclass[aps,prl,reprint,superscriptaddress,showpacs]{revtex4-1}
\usepackage{graphics} 
\usepackage{amsmath}
\usepackage{color}
\begin{document}
 \title{Both electron and hole Dirac cone states in Ba(FeAs)$_2$ confirmed by 
magnetoresistance}
\author{Khuong K. Huynh}\email{khuong@sspns.phys.tohoku.ac.jp}
\affiliation{Department of Physics, Graduate School of
Science,Tohoku University} 
\author{Yoichi Tanabe} \email{youichi@sspns.phys.tohoku.ac.jp}
\affiliation{WPI-AIMR, Tohoku University}
\author{Katsumi Tanigaki}
\email{tanigaki@sspns.phys.tohoku.ac.jp} 
\affiliation{Department of Physics, Graduate
School of Science,Tohoku University}
 \affiliation{WPI-AIMR, Tohoku University} 
\date{\today}

\begin{abstract}
Quantum transport of Dirac cone states in the
iron pnictide Ba(FeAs)$_2$ with a d-\,multiband system is studied by
using single crystal samples.  
Transverse magnetoresistance develops linearly against the magnetic field at low temperatures.
The transport phenomena are interpreted in terms of the 0$^{th}$ Landau level by applying
the theory predicted by Abrikosov. 
The results of the semiclassical analyses of a two carrier system in a low magnetic field limit 
show that both electron and hole reside as the high mobility states.
Our results show pairs of electron- and hole Dirac cone states must be taken into account for accurate interpretation in iron pnictides, which is contrast to previous studies.
\end{abstract}


\maketitle

An intriguing issue in condensed matter physics nowadays is the massless Dirac fermion states in materials, such as graphene, topological insulators, and organic conductors. 
The linear relationship between momentum and energy leads to a very high 
transport mobility due to the zero effective mass and the long relaxation time of the conduction electrons regardless of impurities and/or various many body effects \cite{JPSJ.67.2857}. 
In a magnetic field ($B$), Landau level (LL) splittings of the Dirac cone states are proportional to the square root of the external $B$ strength ($E_n= \pm v_{\text{F}}\sqrt{2\hbar e B |n|}$, where $v_\text{F}$ is the Fermi velocity and $n$ is the LL index). 
Energy scaling makes the LL states thermally stable even in a moderate $B$ (e.g., $B\leq 10 ~\text{T}$) \cite{RevModPhys.81.109}. 
Consequently, the low energy properties of discrete LL states become accessible to conventional experimental probes, especially magneto-transport measurements. 
Recently, intensive research has revealed a variety of interesting quantum magneto-transport phenomena in materials with Dirac cone states, such as Quantum Hall effects (QHE) found in graphene \cite{novoselov2005two} and topological insulators \cite{RevModPhys.82.3045} and unusual magnetoresistance (MR) in the multi-layered $\alpha$-(BEDT-TTF)$_2$I$_3$ organic conductor \cite{PhysRevLett.102.176403}.

Recently, new Dirac cone states have been theoretically predicted
\cite{NewsComments,PhysRevB.79.014505} and experimentally confirmed in Ba(FeAs)$_2$, which is a parent compound of iron pnictide superconductors \cite{kam08,PhysRevB.78.020503},by using angle-resolved photoemission spectroscopy (ARPES) \cite{PhysRevLett.104.137001}.
The present consensus is that iron pnictides are semi-metals and that interband antiferromagnetic (AFM) interactions must be considered in order to understand their intriguing electronic properties, such as relatively high temperature superconductivity and itinerant/localized magnetic properties. 
Ba(FeAs)$_2$ exhibits spin-density-wave (SDW) instability at 138 K, and Fermi surface (FS) nesting eliminates nearly $90~\%$ of the conduction electrons and holes.
Furthermore, this material shows the first Dirac cone states, constructed via
band folding due to d-band AF interactions, in a multiband system.
Theoretical considerations explain that the formation of Dirac cone states is a consequence 
of the nodes of the SDW gap by complex zone foldings in bands with different parities \cite{PhysRevB.79.014505}. 
The Dirac cone states in Ba(FeAs)$_2$ are bulky states induced by 3-dimensional band folding and are different from those of 2-dimensional graphene and the space-inversion symmetry broken surface of topological insulators.
Therefore, dominant contributions and distinguished quantum magneto-transport phenomena from the Dirac cone states can be expected in the transport properties of Ba(FeAs)$_2$.
Thus, it is important to study the Dirac cone states appearing in d-multi band iron pnictides.
Due to moderate electron correlations in a three-dimensional system with a unique multi-band nature in 
a d-electron systems, very different $B$- and $T$- dependences in a discrete LL will be observed from  the Dirac cone states of Ba(FeAs)$_2$. 

In this letter, we have reported the first investigation on MR of the Dirac cone states in Ba(FeAs)$_2$. 
We have investigated the $B$ and $T$ dependence of the in-plane MR and Hall coefficient ($R_\text{H}$) of single crystalline samples. 
The observed dependence of MR on $B$ is linear in a $B$ of moderate strength ($|B|\geq2~\text{T}$), but changed from a linear to a squared relation (MR $ \approx a B^2$) in low $B$s ($|B|\leq0.5 ~\text{T}$). 
We have interpreted the linear MR in $B$ of moderate strength as the inherent quantum limit of the $0^\text{th}$ LL of the Dirac cone states in accordance with Abrikosov's model of a quantum MR \cite{PhysRevB.58.2788}. 
Since the semiclassical transport phenomena is observed when $|B|\leq0.5 ~\text{T}$, important physical parameters can be estimated from the low $B$ data using semiclassical two-carrier-type analyses.
The mobilities of both electrons and holes are 8.7 times greater at the SDW transition. 
Contributions from pairs of electron- and hole-like Dirac cone states to the transport properties of iron pnictide materials are reported.

Single crystals of Ba(FeAs)$_2$ were grown by employing an FeAs self-flux method described in detail elsewhere \cite{PhysRevB.78.214515}. 
The crystal quality was confirmed by using synchrotron X-ray diffraction at the beamline BL02B2 in SPring-8. 
Studies on the $T$- and $B$-dependences of the in-plane transverse MRs and $R_\text{H}$s were carried out using a Quantum Design physical property measurement system (PPMS) in a $B$ range of -9 to 9 T in the range of 2 - 300 K. 
The electric current was in the ab plane and the applied $B$ was parallel to the c-axis.

\begin{figure} 
	\scalebox{0.25}{\includegraphics{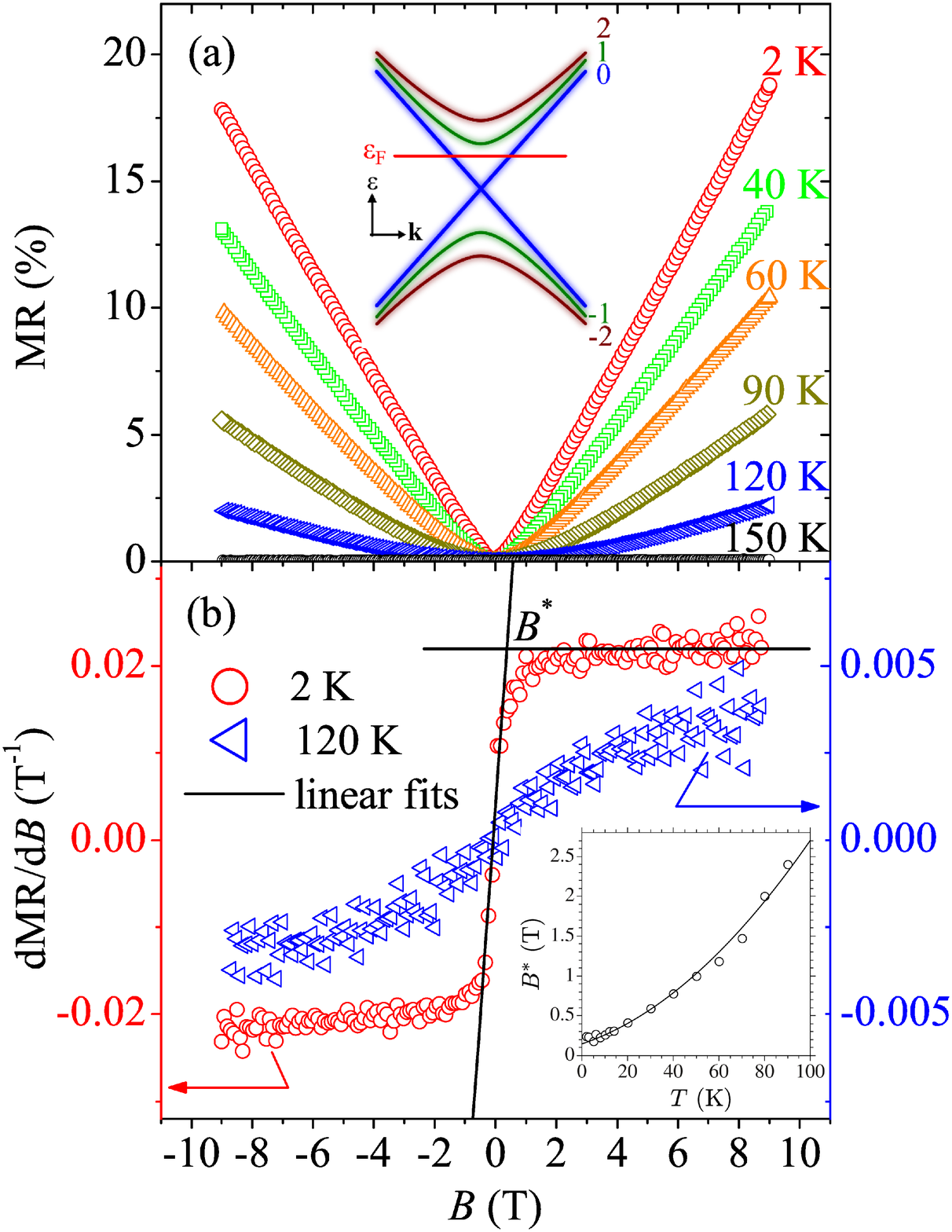}} \qquad 
	\caption{(Color online) (a) The $B$-dependence of MR in the temperatures ($T$) range of 2 - 150 K. 
The inset of Fig. 2(a) illustrates the idea of Abrikosov model, in which all of the carriers are confined in the $0^\text{th}$ LL. 
The colored numbers denote the indexes of the LLs and the shadings depict the thermal broadening of the LLs. 
(b) The magnetic field ($B$) derivatives of MR at 2 K and 120 K. 
The thin black line denotes the semiclassical approximation of MR$\approx B$
at $2~\text{K}$. 
The crossover of the $B^*$ between the semiclassical regime and the quantum linear regime is marked with the black arrows. 
The inset of Fig. 2(b) shows the temperature dependence of $B^*$ (black circles) up to 50 K. 
The black solid line was fitted using $B^*(T)=\left(1/2e \hbar v_\text{F} ^2 \right) \left (E_F + k_\text{B} T \right )^2$.} 
	\label{fig1}
\end{figure}

Figure \ref{fig1}(a) shows a typical $B$ dependence of the in-plane MR for Ba(FeAs)$_2$ in the temperature range of 2 - 150 K, where $\text{MR}(B)=\left[\rho(B)-\rho(0)\right]/\rho(0)$. 
Above the SDW transiton at $T^*=138 \text{K}$, $\text{MR}\leq~0.02\%$ in the entire $B$ range without any clear dependence on $B$. 
Below $T^*$, the value of MR increased with a decrease in the temperature with an unusual linear symmetrical V-shape curve in the high $B$ regime, whereas a small parabolic-like bend remains in the low $B$ limit. 
This is in sharp contrast to the behavior of other semi-metals, in which MR generally develops with a dependence on $B^2$ over the entire range of $B$. 
The linear dependence on $B$ is evident when the first-order derivative $d\text{MR}(B)/dB$ curve is examined, as shown in Fig.\,\ref{fig1}b.
$d\text{MR}(B)/dB$ saturated in a large $B$.
In a low $B$ (e.g., $|B| < 0.5~ \text{T}$ at 2 K), $d\text{MR}(B)/dB$ crossovers to a semiclassical $B$-dependence where $d\text{MR}/dB$ is proportional to B. 
The critical $B^*$, defined as the intersection between the straight line and the horizontal line, where the $B$-linear MR is observed in high $B$-regime, is considered to be the crossing point between the semiclassical and the quantum linear transport regimes. 
At low $T$, the linear V-shape dependence of MR became more pronounced, whereas the crossover behavior gradually smeared out as $T$ approached $T^*$, as illustrated in Fig.\,\ref{fig1}(b). 
The observed linear $B$-dependent MR clearly represents a magneto-transport property of the Dirac cone states, which will be discussed later.

\begin{figure} 
		\scalebox{0.33}{\includegraphics{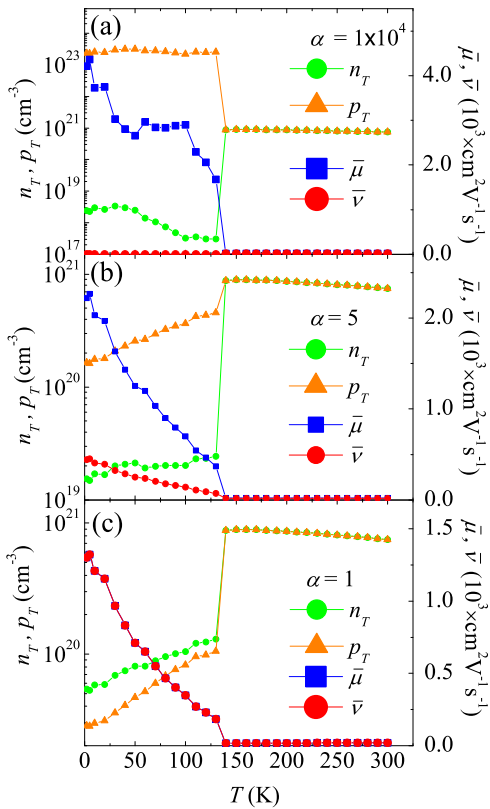}} \qquad 
		\caption{(Color online) Total numbers and averaged mobilities of electrons and holes obtained from semiclassical analyses with (a) $\alpha=1\times10^4$, (b) $\alpha=5$, and (c) $\alpha=1$, which were determined assuming one-carrier approximation, two carrier system dominated by fast electrons, and semi-metals, respectively.}\label{fig2} 
	\end{figure}

When $|B|<B^*$, both $R_\text{H}$ and MR obeyed a two-carrier-type semiclassical model, i.e., $\text{MR}(B)$ is proportional to $B^2$ and $R_\text{H}$ is nearly $\text{constant}$. 
One can therefore utilize the datasets in this $B$ range to estimate transport parameters, such as the numbers and the mobilities of the carriers. The zero-field resistivity ($\rho(0)$), MR, and $R_\text{H}$ in the low $B$ limit can be written as  
	\begin{eqnarray}
		\rho(0)&=&\displaystyle{\frac{1}{e\left(n_{\rm T}\bar{\mu}+p_{\rm T}\bar{\nu}\right)}}, \label{1}\\
		\text{MR}&=&\displaystyle{\frac{\rho(B)-\rho(0)}{\rho(0)}=\frac{n_{\rm T} p_{\rm T} \bar{\mu} \bar{\nu}\left(\bar{\mu}+\bar{\nu}\right)^2}{\left(n_{\rm T}\bar{\mu}+p_{\rm T} \bar{\nu}\right)^2}B^2}, \label{2}\\
		R_\text{H}&=&\displaystyle{\frac{-n_{\rm T} {\bar{\mu}}^2+p_{\rm T} {\bar{\nu}}^2}{e\left(n_{\rm T}\bar{\mu}+p_{\rm T}\bar{\nu}\right)^2}}.  \label{3}
	\end{eqnarray}
Here $n_{\rm T} \equiv \left( n_{\rm P}+n_{\rm D}\right)$ and $p_{\rm T}\equiv \left( p_{\rm P}+p_{\rm D}\right)$ are the total numbers of electrons and holes, and $\bar{\mu} \equiv \left(n_{\rm P} \mu_{\rm P}+n_{\rm D}\mu_{\rm D}\right)/n_{\rm T}$ and $\bar{\nu}\equiv \left(p_{\rm P} \nu_{\rm P}+p_{\rm D}\nu_{\rm D}\right)/p_{\rm T}$ are the averaged  mobilities of electrons and holes, respectively. 
The subscript $``D"$ and $``P"$ denote the Dirac cone states and the parabolic bands, respectively. 
In order to estimate $n_{\rm T}, p_{\rm T}, \bar{\mu}$, and $\bar{\nu}$ from Eqs.~(\ref{1})-(\ref{3}), an additional condition is necessary. 
Above $T^*$,  Ba(FeAs)$_2$ is a semi-metal without Dirac cone states, i.e., $\kappa\equiv n_{\rm P}/p_{\rm P} \cong 1$ and $n_{\rm D}=p_{\rm D} \equiv 0 $ as determined by using ARPES \cite{PhysRevB.80.024515}. 
Therefore, we analyzed the data using the condition $\kappa=n_{\rm P}/p_{\rm P}=1$ \footnote{Since the $\text{MR}\leq0.02~\%$ when $T>T^*$, the value $0.02~\%$ was used as an upper limit to estimate the parameters.}. 
In the SDW state below $T^*$, where the Dirac cone states coexist with several electron and hole pockets of parabolic bands \cite{PhysRevB.80.174510}, we analyzed the data using various values of the mobility ratio $\alpha \equiv \bar{\mu}/\bar{\nu}$. 
The estimated transport parameters with $\alpha$ = 1 $\times$ 10$^4$, 5, and 1 are shown in Figs.\,\ref{fig2}(a)-(c). 

MRs linearly dependent on $B$ have been reported for several materials, such as Bi and InSb \cite{yang1999large}. 
Abrikosov interpreted this phenomenon by considering a quantum limit where all of the carriers in the Dirac cone states occupy only the $0^{\text{th}}$ LL, as depicted in the inset of Fig.\,\ref{1}(a) \cite{PhysRevB.58.2788}. 
This situation can be realized when the following two specific conditions are taken into account. 
First, the LL splitting of $\Delta_1=|E_{\pm1}-E_0|=\pm v_{\text{F}}\sqrt{2\hbar e B} $ between the $1^\text{st}$ and the $0^\text{th}$ LLs must be larger than the Fermi energy $E_\text{F}$ of the system. 
This means that all carriers can occupy only the $0^\text{th}$ LL and that $B$ is higher than the critical value $B^*(0)$ at $0$ K. 
Second, the thermal fluctuation at a finite temperature ($k_{\text{B}}T$) does not exceed $\Delta_1$. 
In such a quantum limit, MR can no longer be described within the framework of the conventional Born scattering approximation, and is instead expressed by $\text{MR}\propto \left(N_i / en_{D}^2\right)B$, where the Dirac carriers $n_{\rm D}$ and the impurities $N_i$ determine the MR. 
The resulting MR is therefore linear in relation to $B$.

In a parabolic band, the LL splitting is proportional to the first order of $B$, i.e. $\Delta_n=\hbar eB/m^*$, where $m^*$ is the effective mass defined by the curvature of the band. 
Thus, a huge $B$ strength is needed in order to satisfy the condition $\Delta_n>k_{\rm B}T$.
In contrast, the LL splitting $\Delta_1$ in a Dirac cone state scales with the square root of $B$, leading to a much larger LL splitting than that in a parabolic band under the same $B$ strength. 
Consequently, the quantum limit of Dirac cone states can be realized in a moderate $B$ \cite{PhysRevB.58.2788}.
The $B$-linear MR when $B<9~\text{T}$, therefore, originates from the Dirac cone states.  
The observed $B^*(T)$ corresponds to the limit of 
$B^*=\left(1/2 e\hbar v_\text{F}^2\right)\left(k_\text{B} T + E_\text{F} \right)^2$, at which $\Delta_1$ = $E_{\rm F}$ + $k_\text{B} T$.
Hence, the observed MR changes from the quantum limit to the semiclassical one when $|B|<B^*$.
The inset of Fig.\,\ref{1}(b) shows values of $B^*$ at different temperatures up to $100~\text{K}$ together with a curve fitted by using $B^*(T)=\left(1/2 e\hbar v_\text{F}^2\right)\left(k_\text{B} T + E_\text{F} \right)^2$. 
A good agreement of $B^*(T)$ with the above equation confirms the role of Dirac cone states in the observed $B$-linear MR.
The crossover point between the semiclassical and the quantum transport regimes gradually smeard out when $T$ is in the vicinity of $T^*$ as illustrated in Fig.\,\ref{fig1}(b), since the Dirac cone states are the nodes of the SDW gap and are associated with the $T$-evolution of the SDW order parameter \cite{PhysRevLett.104.137001}.

As a result of the analyses, we obtained $v_{\text{F}} \approx 1.88 \times 10^5~\text{ms}^{-1}$ and 
$B^*(0) \approx 0.15~\text{T}$.
The value of $B^*(0)$ corresponds to $\Delta_1 \approx 2.48~\text{meV}$, which is almost consistent with $E_\text{F} = 1 \pm 5 ~\text{meV}$ reported by \cite{PhysRevLett.104.137001}.
This value of $\Delta_1$ indicates that only the $0^\text{th}$ LL exists below $E_\text{F}$ 
when $B$ $\geq$ $B^*$. 
LDA calculations have shown that the electronic structure of Ba(FeAs)$_2$ is composed of two large parabolic-like and two small Dirac-like FSs with different sizes \cite{PhysRevB.80.064507}. 
We attributed the linear MR observed in our experiments to the smallest Dirac cone state. 
Shubnikov-de Haas (ShdH) oscillations can arise from either parabolic FSs or other Dirac cone 
states with larger $E_\text{F}$. 
When $|B|\leq 9~\text{T}$, the LL splitting of the parabolic band is too small to observe ShdH oscillations. 
On the other hand, ShdH oscillations of the larger Dirac cone state is possible 
owing to the different energy scale $\Delta_1=v_\text{F}\sqrt{2\hbar e B}$.
This can be the next intriguing topic for studying quantum behaviors of the Dirac cone 
states in Ba(FeAs)$_2$.

Although the linear $B$-dependence of MR is clear evidence for Dirac cone states, a question still remains as to whether there is only a single electron-like Dirac cone state or theree are pairs of electron-like and hole-like Dirac cone states in the SDW state of Ba(FeAs)$_2$ \cite{PhysRevB.80.064507,PhysRevB.80.224512}.
From ARPES data \cite{PhysRevB.80.174510}, the Luttinger volumes of electron and hole pockets projected into the $(k_x,k_y)$ Brillouin zone are comparable, i.e, $\kappa=n_T/p_T\cong 1$ in the SDW state.
As shown in Figs.\,\ref{fig2}(a)-(c), the value of $\kappa$ reflecting the ARPES data was only reproduced by using the semiclassical analysis with $\alpha=1$ \footnote{One can choose to fix $\kappa$ at some values near 1 and let $\bar{\mu}$ and $\bar{\nu}$ be free parameters. Such analyses yield almost the identical results with that in the main text, i.e. $\alpha=\bar{\mu}/\bar{\nu}\cong 1$}. 
As shown in Fig.\,\ref{2}(c), both $n_T$ and $p_T$ were around $7.0\times10^{20}~\text{cm}^{-3}$ with $\bar\mu$ and $\bar\nu$ around $25~\text{cm}^2\text{V}^{-1}\text{s}^{-1}$ above $T^*$, being in consistent with the medium electronic correlations observed in iron pnictides \cite{qazilbash2009electronic}.
Below $T^*$, both $n_T$ and $p_T$ decreased by $90~\%$ due to the SDW nesting. 
At the same time, both $\bar\mu$ and $\bar\nu$ jumped by $8.7$ times, indicating the existence of high mobility Dirac carriers in the SDW state. 
More significantly, the simultaneous increase in $\bar\mu$ and $\bar\nu$ directly demonstrates the coexistence of electron- and hole-like Dirac cone states, although the contribution of the electron and hole pockets in parabolic bands, remaining still after the SDW transition, cannot be ignored.  
Our conclusions here are in strong contrast with the previous arguments suggesting that the high mobility of the electrons play a dominant role in transport properties \cite{PhysRevB.81.020510,PhysRevB.80.140508,PhysRevLett.103.057001,PhysRevLett.105.037203}.

Our present results have unveiled the interesting relationship between the transport properties and the electronic states of Ba(FeAs)$_2$. 
Below $T^*$, electron- and hole-like parabolic bands coexist with electron-like and hole-like Dirac cone states formed at different $\bf{k}$-points. 
The two Dirac cone states control the magneto-transport phenomena over the entire $B$ range.
The pairs of electron- and hole-like Dirac cone states observed in the present experiments agree with LDA band calculations and dHvA experiments \cite{PhysRevB.80.064507,PhysRevB.80.224512}.
A hole-like Dirac cone state has been observed in the $k_z=0$ plane by ARPES, and no other Dirac cone states cannot be observed in the same  $k_z$ plane.
In order to fully interpret our experimental observations and the previous reports, another electron-like Dirac cone state must exist at a $k_z\neq 0$ position.
Experimental approaches to find such a state are warranted. 
The intrinsic transport properties of Ba(FeAs)$_2$ have contributions from the carriers residing both in parabolic bands and in Dirac cone states. 
The former inherently have low mobilities due to electronic correlations, whereas the latter have high mobilities in spite of the many-body, even by impurity scattering.
This explains the sharp increase in the conductivity observed at the SDW transition, although almost all of the carriers disappear at the transition temperature \cite{NewsComments}.

In summary, we observed a $B$-linear MR, exhibiting quantum transport in the $0^\text{th}$ LL of Dirac cone states of Ba(FeAs)$_2$. 
To the best of our knowledge, this is the first evidence of quantum transport involving the Dirac cone states in iron pnictides. 
Four important transport parameters, i.e., the total numbers and the mobilities of electrons and holes, were successfully evaluated using $\rho(0), \text{MR, and}~R_\text{H}$. 
The present results showed that electron- and hole-like Dirac cone states existed in pairs in iron pnictides. 
The superconductivity appearing in the iron pnictide family is still a hot issue, and magnetic interactions are thought to be very important. 
The superconducting dome eventually evolves before the AFM ordering state disappears, and therefore, the relation between the Dirac cone states and the superconductivity in the vicinity of the quantum critical point must be investigated. 

\begin{acknowledgments} 
We acknowledge H. Fukuyama, T. Takahashi, F. Sato, S. Souma, E. Kaneshita, N. Kimura, B. Breedlove, J. Tang, and G. Mu for their discussions. The research was supported by a Grant-in-Aid for Scientific Research on Priority Areas of New Materials Science using Regulated Nano Spaces from the Ministry of Education, Science, Sports, Culture and Technology of Japan. The work was partly supported by the Tohoku GCOE Program and by the Japan Synchrotron Radiation Research Institute (JASRI). 
\end{acknowledgments}

\textit{Note\,added\,in\,proof} 
During the referee process, a report of magnetotransport measurements in Ba(Fe$_{1-x}$Co$_x$As)$_2$ employing the same semiclassical analyses as those in the present paper was submitted to Cond.Mat. (arXiv:1103.4535, March 23$^\text{rd}$, 2011).

%
\end{document}